\documentclass[twocolumn]{aastex6}
\defcitealias{gabanyi2013}{GDG13}
\defcitealias{tanaka2014}{TSF14}
\defcitealias{liuzzo2013}{LGG13}
\slugcomment{A manuscript resubmitted to \apjl}
\shorttitle{EVN Observations of HESS J1943+213}
\shortauthors{K. Akiyama et al.}
\usepackage{color}

\newcommand{\haystack}{1}
\newcommand{\naoj}{2}
\newcommand{\utokyo}{3}
\newcommand{\uhiroshima}{5}
\newcommand{\ujagiellonian}{4}
\newcommand{\inaf}{6}
\newcommand{\sokendai}{7}
\begin{document}
	\title{EVN Observations of HESS J1943+213: Evidence for an Extreme TeV BL Lac Object}
	
	\author{Kazunori Akiyama\altaffilmark{\haystack,\naoj,\utokyo,8}}
	\author{{\L}ukasz Stawarz\altaffilmark{\ujagiellonian}}
	\author{Yasuyuki T. Tanaka\altaffilmark{\uhiroshima}}
	\author{Hiroshi Nagai\altaffilmark{\naoj}}
	\author{Marcello Giroletti\altaffilmark{\inaf}}
	\author{Mareki Honma\altaffilmark{\naoj,\sokendai}}
	%
	%
	\altaffiltext{\haystack}{Massachusetts Institute of Technology, Haystack Observatory, Route 40, Westford, MA 01886, USA}
	\altaffiltext{\naoj}{National Astronomical Observatory of Japan, 2-21-1 Osawa, Mitaka, Tokyo 181-8588, Japan}
	\altaffiltext{\utokyo}{Department of Astronomy, Graduate School of Science, The University of Tokyo, 7-3-1 Hongo, Bunkyo-ku, Tokyo 113-0033, Japan}
	\altaffiltext{\ujagiellonian}{Astronomical Observatory, Jagiellonian University, ul. Orla 171, 30-244 Krak\'ow, Poland}
	\altaffiltext{\uhiroshima}{Hiroshima Astrophysical Science Center, Hiroshima University, 1-3-1 Kagamiyama, Higashi-Hiroshima 739-8526, Japan}
	\altaffiltext{\inaf}{INAF Istituto di Radioastronomia, via Gobetti 101, 40129 Bologna, Italy}
	\altaffiltext{\sokendai}{Graduate University for Advanced Studies, Mitaka, 2-21-1 Osawa, Mitaka, Tokyo 181-8588}
	%
	\altaffiltext{8}{\url{kazu@haystack.mit.edu}; JSPS Postdoctoral Fellow for Research Abroad}
	%
	\begin{abstract}
		We report on the 1.6~GHz (18~cm) VLBI observations of the unresolved, steady TeV source HESS~J1943+213 located in the Galactic plane, performed with the European VLBI Network (EVN) in 2014. Our new observations with a nearly full EVN array provide the deepest image of HESS J1943+213 at the highest resolution ever achieved, enabling us to resolve the long-standing issues of the source identification. The milliarcsecond-scale structure of HESS~J1943+213 has a clear asymmetric morphology, consisting of a compact core and a diffuse jet-like tail. This is broadly consistent with the previous e-EVN observations of the source performed in 2011, and re-analyzed in this work. The core component is characterized by the brightness temperature of $\gtrsim1.8 \times 10^9$~K, which is typical for low-luminosity blazars in general. Overall, radio properties of HESS~J1943+213 are consistent with the source classification as an ``extreme high-frequency-peaked BL Lac object''. Remarkably, we note that since HESS~J1943+213 does not reveal any optical or infrared signatures of the AGN activity, it would never be recognized and identified as a BL Lac object, if not its location close to the Galactic plane where the High~Energy~Stereoscopic~System has surveyed, and the follow-up dedicated X-ray and radio studies triggered by the source detection in the TeV range. Our results suggest therefore a presence of an unrecognized, possibly very numerous population of particularly extreme HBLs, and simultaneously demonstrate that the low-frequency VLBI observations with high-angular resolution are indispensable for a proper identification of such objects.
	\end{abstract}
	%
	%
	\section{Introduction}
	A variety of very-high-energy (VHE; $E>100$~GeV) $\rm \gamma$-ray sources have been discovered in the Galactic plane by recent systematic surveys with the High Energy Stereoscopic System \citep[H.E.S.S.;][]{carrigan2013}. The majority of these sources are Galactic objects with extended structures such as supernova remnants (SNRs), evolved pulsar wind nebulae (PWNe), or molecular clouds. Unresolved or point-like H.E.S.S. sources in the Galactic plane have been identified with Galactic high-mass X-ray binaries (HMXBs), or young PWNe (e.g., Crab Nebula). On the other hand, outside of the Galactic plane, the vast majority of the unresolved $\rm \gamma$-ray sources are radio-loud active galactic nuclei (AGNs). More than 50 AGNs have been detected in VHE $\gamma$-ray regime so far, which are mostly blazars of the BL Lacertae type (hereafter BL Lacs)\footnote{\url{http://tevcat.uchicago.edu}}.
	
	The unresolved, steady TeV source HESS~J1943+213 (henceforth J1943+213) was discovered in the Galactic plane with three-years H.E.S.S. observations between 2005 and 2008 \citep{abramowski2011}. The nature of J1943+213 is puzzling and subjected to an ongoing debate. It is located within the $4.4^{\prime}$ error circle of an unidentified hard X-ray {\it INTEGRAL} source IGR J19443+2117, which was also seen with {\it ROSAT}, {\it Chandra} and {\it Swift} \citep{tomsick2009,landi2009,cusumano2010}. \citet{abramowski2011} discussed in detail the AGN and PWN identification of J1943+213 and, based on the gathered multi-wavelength data, concluded that the most likely classification of the source is that of an ``extreme high-frequency-peaked BL Lac object'' (``extreme HBL'' for short). 
	
	On the other hand, following-up radio observations with optically-connected stations in European VLBI Network (e-EVN mode) revealed a compact radio counterpart to the TeV emitter, J1943+2113 \citep[][henceforth \citetalias{gabanyi2013}]{gabanyi2013}.  \citetalias{gabanyi2013} reported non-detection of signals on the longest baselines between European stations and the African station at Hartebeesthoek (Hh), indicative of a source extension on milliarcsecond-scale. This, along with the derived low brightness temperature, argued against the blazar identification. \cite{leahy2012} reported the HI absorption spectrum indicating the source distance exceeding 16~kpc. However, it is not conclusive evidence for its extragalactic origin, since it can still be a PWN at a distance of  $\sim$17~kpc \citepalias{gabanyi2013}.
	
	Recently, new X-ray observations of J1943+213 with {\it Suzaku} have been reported by \citet[][hereafter \citetalias{tanaka2014}]{tanaka2014}, who re-analyzed also the infrared data for the source from the WISE and UKIDSS surveys, as well as the {\it Fermi} Large Area Telescope ({\it Fermi}/LAT) data integrated over 4.5~years (yielding the GeV flux upper limits improved over those derived in \citealt{abramowski2011}). The best-quality X-ray spectrum revealed a single power-law continuum extending up to 25~keV, with a moderate absorption in excess of the Galactic value. Moreover, the re-analyzed infrared data have been found to be consistent with the presence of a luminous elliptical host located at the luminosity distance of $\sim 600$~Mpc. All these findings, together with the broad-band modeling of the source spectral energy distribution (SED) performed by \citetalias{tanaka2014} in the framework of the extreme HBL scenario, supported the blazar identification of J1943+213. This was further substantiated by \cite{peter2014}, who carried-out near-infrared imaging of the host candidate with the CAHA telescope, and also reported the source detection in the accumulated five-year-long {\it Fermi}/LAT dataset. Note that \citet{straal2016} have recently reported on the Arecibo observations failing to uncover the putative pulsar powering the PWN, again supporting the extreme HBL scenario.
	
	\begin{figure}[t]
		\centering
		\includegraphics[width=1\columnwidth]{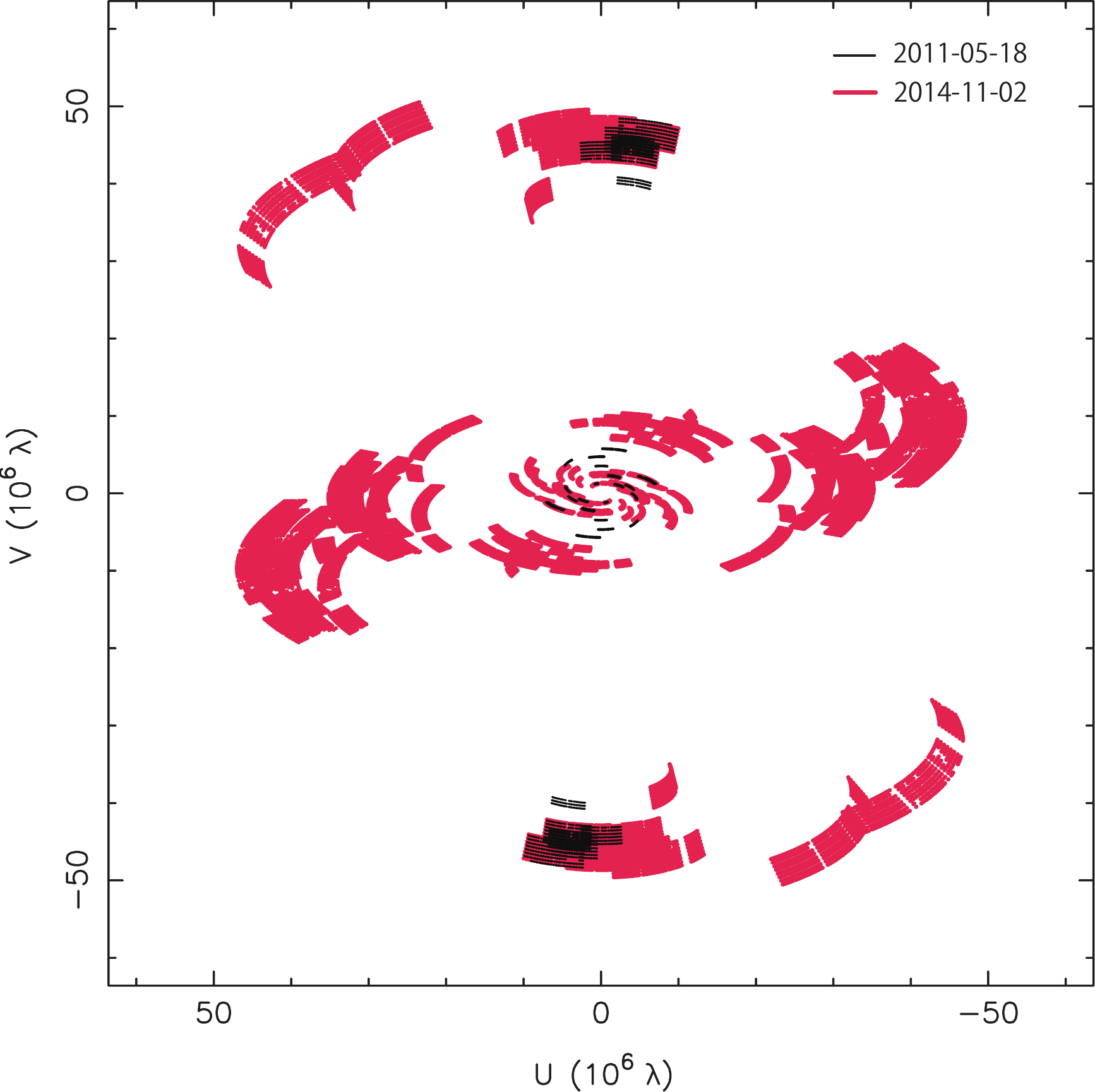}
		\caption{
			The $uv-$coverages of EVN observations. The black curves indicate the $uv-$coverage of 2011 observations with e-EVN, while the red curves correspond to the $uv-$coverage of our new EVN observations in 2014.
			\label{fig:uv-coverage}}
	\end{figure}

	The small-scale radio structure of J1943+213 is the remaining key puzzle in the discussion regarding the nature of the source. Although the previous high-sensitivity e-EVN observations have enabled to successfully detect and locate the radio counterpart to J1943+213, the corresponding spatial resolution and imaging sensitivity were rather limited by poor $uv-$coverages due to the lack of other EVN stations and hour-angle coverages. More sensitive observations at higher angular resolution are therefore indispensable for a proper characterization of the source at radio frequencies.
	
	In this Letter, we report on the new EVN observations of J1943+213, combined with the re-analysis of the e-EVN data taken in 2011 (from \citetalias{gabanyi2013}). Our new EVN observations with nearly full array provide the deepest image of the radio structure of J1943+213 on milliarcsecond scales at the highest achievable resolution.
	
	\section{Observations and Data Reductions}
	\subsection{2011 observations}
	We reduced the archival e-EVN data for J1943+213 at 1.6~GHz, which were presented before in \citetalias{gabanyi2013}. Observations were performed with seven telescopes, as summarized in Table~\ref{tab:obs-sum}, at a recording rate of 1024~Mbit~s$^{-1}$ on May 2011 for $\sim 2$~h. The total bandwidth was 128~MHz per polarization.
	
	Initial calibrations were performed in the Astronomical Image Processing System (AIPS). The visibility amplitudes were calibrated in the AIPS task {\tt APCAL} based on the system temperature and aperture efficiency measured at each station. 
		
	We found that non-detections of fringes on the longest baselines from European telescopes to the Hh station in Africa reported in \citetalias{gabanyi2013} are not owing to the extended structure. \citetalias{gabanyi2013} performed phase-reference VLBI observations towards J1943+213, since the source position was highly uncertain, and calibrated the data with the EVN pipeline\footnote{\url{http://www.evlbi.org/pipeline/user\_expts.html}}. In the pipeline processing, the visibility phase was calibrated with the nearby phase calibrator J1946+2300.
	We found that no fringes were detected on Hh baselines for J1946+2300 in the pipeline calibrations, because the source structure is resolved out in long baselines. This would be the reason why \citetalias{gabanyi2013} missed the fringes also in the case of J1943+213, where the fringe-search solutions of J1946+2300 were applied.

	\begin{figure}[t]
		\centering
		\includegraphics[width=1\columnwidth]{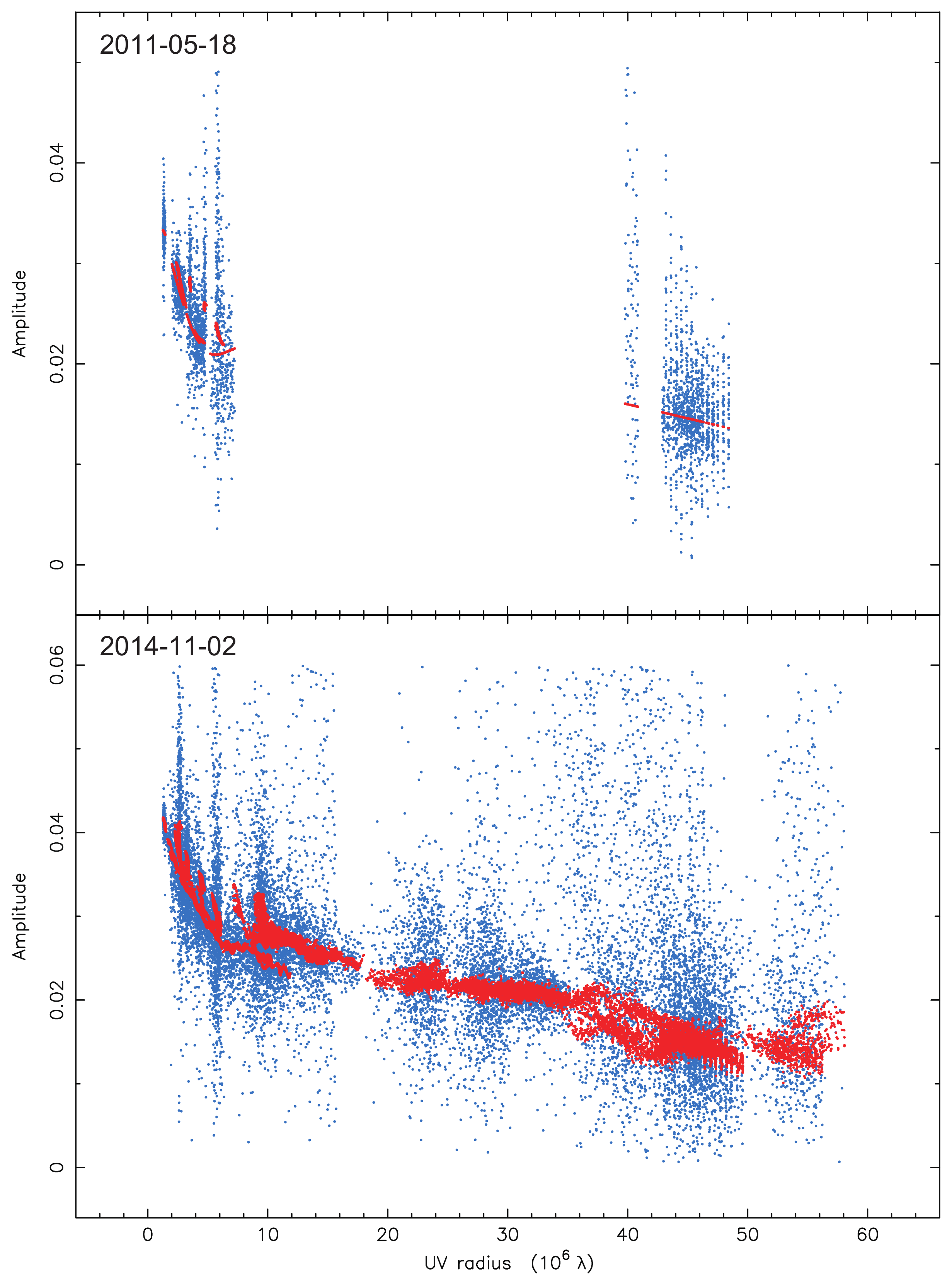}
		\caption{
			The calibrated visibility amplitude distribution as a function of the baseline length, obtained with e-EVN observations in 2011 (top panel) and our new EVN observations in 2014 (bottom panel). The blue points indicate the visibility amplitudes, while the red points indicate the model visibility amplitudes of the best-fit Gaussian components for 2011 observations and the CLEAN components for 2014 observations. Note that both data are further time averaged for the integration time of 120~s for reducing number of points. The unit of the vertical axis is Jy. 
			\label{fig:amp}}
	\end{figure}

		\begin{table*}[t]
		\centering
		\caption{Summary of EVN observations and images\label{tab:obs-sum}}
		\begin{tabular}{ccclcccc}
			\hline \hline
			Date & Frequency & Obs Code & Stations & Beam Size & Beam PA & Peak Intensity & Image rms\\
			(yyyy-dd-mm) & (GHz) & & & (mas) & ($^\circ$) & (mJy~beam$^{-1}$)& ($\mu$Jy~beam$^{-1}$)\\
			\hline
			2011-05-18 & 1.664 & RSG03 & Jb Wb Ef On Tr Mc Hh & 33.1$\times$2.03 & 83.9 & 18.2 & 64\\
			2014-11-02 & 1.658 & EA056B & Jb Wb Ef On Sh Tr Sv Zc Bd Hh & 3.48$\times2.93$ & 85.4 & 19.9 & 12 \\
			\hline
		\end{tabular}
	\end{table*}
	
	In this work, fringe fitting was directly applied to J1943+213 in the AIPS task {\tt FRING}, after removing phase, delay and rate offsets due to the source positional error by correcting its coordinate to $(\alpha,\delta)=(19^h43^m56^s.2372,21^\circ18'23''.402)$ measured in \citetalias{gabanyi2013} with the AIPS task {\tt CLCOR}. We successfully detected fringes directly even on the Hh baselines with reliable signal-to-noise ratios (S/Ns).

	Data were averaged for the integration time of 30~s, and then self-calibrated in the package Difmap. The resultant uv-coverage is shown in Fig.~\ref{fig:uv-coverage}. The imaging fidelity and spatial resolution are limited by the relatively poor $uv-$coverage, due to the lack of intermediate/long baselines in the east-west direction and short hour-angle coverages (see \S3 for details). Thus, we conducted following-up observations with EVN at 1.6~GHz in 2014 as described in the next subsection. 
	
	\subsection{2014  observations}
	We conducted EVN observations at 1.6~GHz in November 2014. Observations were performed with 10 telescopes summarized in Table~\ref{tab:obs-sum} at a recording rate of 1024~Mbit~s$^{-1}$ for $\sim4$~h. The total bandwidth was 128~MHz per polarization.
	
	We used pipeline-calibrated data sets. In the pipeline calibration, the visibility amplitudes were calibrated in the AIPS task {\tt APCAL} based on the system temperature and aperture efficiency measured at each station. Fringe fitting was directly applied to J1943+213 in the AIPS task {\tt FRING}. The fringe was robustly detected in all the stations with reliable S/Ns, consistently with the 2011 observations.
	
	Data were averaged for the integration time of 30~s, and self-calibrated in Difmap. The resultant $uv-$coverage, shown in Fig. \ref{fig:uv-coverage}, was significantly improved by involving Russian and East Asian stations, and also by a wider hour-angle coverages.
	
	\section{Results}
	\subsection{2011 observations}
	\label{sec:3.2}
	We have successfully detected fringes on long baselines between Europe and the Hh station in Africa by reducing the data manually. The correlated flux density is typically $\sim 12-14$~mJy for these baselines (Fig.~\ref{fig:amp}), which is indeed brighter than the typical baseline sensitivity of $7\sigma _{\rm rms} \sim$~6.6~mJy between the Effelsberg station and the Hh station\footnote{\url{http://www.evlbi.org/user\_guide/base\_sens.html}}. Note that the newly detected correlated flux density is much larger than that following from the best-fit single Gaussian model with a FWHM size of 15.8~mas (as reported in \citetalias{gabanyi2013}), and hence J1943+213 appears more compact in the re-analyzed dataset.
	
	Despite a rather limited $uv-$coverage (see Fig.~\ref{fig:uv-coverage} and \ref{fig:amp}), the visibility of J1943+213 can well be modeled with the DIFMAP command {\tt modelfit} assuming two circular Gaussians components with different sizes. The best-fit model image is shown in Fig.~\ref{fig:image}, and its fit to the visibility amplitude in Fig.~\ref{fig:amp}. The model image has a core--jet-like structure elongated in the north-west direction. The total flux density is $34$~mJy, which is broadly consistent with $31$~mJy reported in \citetalias{gabanyi2013}.
	
	\footnotetext[4]{Note that the errors are purely statistical and then do not include the systematic errors caused by residual calibration gains that are practically difficult to be estimated only with single-epoch observations. Hence, the realistic uncertainties could be larger than estimated values, although it does not affect results and discussions of this Letter.}

	\subsection{2014 observations}
	Fringes were successfully detected on all the stations throughout the whole observational time. The correlated flux density is typically $\sim 15$ mJy on long baselines to the Hh station (see Fig.~\ref{fig:amp}) in a good agreement with the 2011 observations.
	
\begin{figure*}[t]
		\centering
		\includegraphics[width=0.8\textwidth]{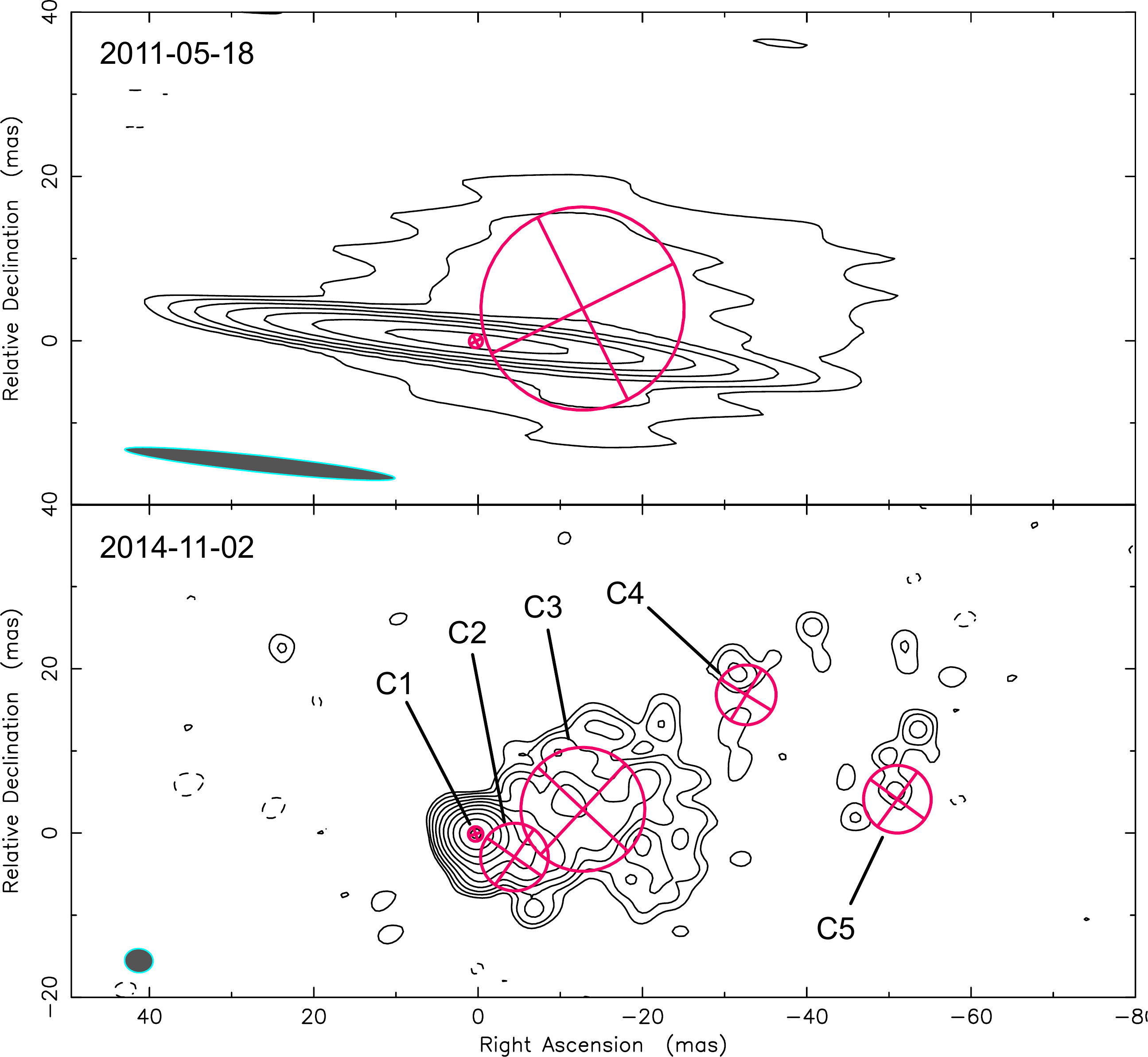}
		\caption{
			The radio images of J1943+213 obtained with the EVN observations. The contour levels are $3 I_{\rm rms} \times (-1,1,2,4,8,...)$, where $ I_{\rm rms}  $ is the image rms of each image. The spatial resolution, peak intensity, and image rms are given in Table~\ref{tab:obs-sum}. {\it Top panel:} the best-fit model image obtained from the e-EVN observations in 2011. The model circular Gaussian components are denoted by magenta circles. {\it Bottom panel:} the uniform-weighted CLEAN image obtained from 2014 observations with EVN. The best-fit circular Gaussian models (Table~\ref{tab:cgauss}) are also shown with magenta circles. 
			\label{fig:image}}
	\end{figure*}

	We show the uniform-weighted CLEAN image in Fig.~\ref{fig:image}. The image clearly reveal a one-side core--jet morphology typical of blazars, with a compact core and a diffuse emission extending to the west/north-west from the core, consistently with the re-analyzed 2011 e-EVN data. The total flux density of the source is 42~mJy. The source structure can well be modeled with five circular Gaussian components, for which the corresponding parameters are given in Table~\ref{tab:cgauss}; the best-fit model is also shown in Fig.~\ref{fig:image}. The positions and sizes of the core component (C1) and the brightest jet feature (C3) are broadly consistent with those resulting from the two-component model applied to the re-analyzed 2011 data. Note that we derived estimates of the 1$\sigma$ uncertainties of each parameter from one-third of its $3\sigma$ confidence interval using the percentile Bootstrap method \citep[see][]{akiyama2013} with $10^4$ data-sets re-sampled from observational data\footnotemark.
		
\begin{table*}[t]
	\centering
	\caption{Circular Gaussian fits to 2014 data with statistical 1$\sigma$ errors.\label{tab:cgauss}}
	\begin{tabular}{cccccc}
		\hline
		\hline 
		ID & Total Flux & \multicolumn{2}{c}{Position} & FWHM & Brightness \\
		& Density & RA & Dec & size & Temperture\\
		& (mJy) & (mas) & (mas) & (mas) & (K)\\
		\hline 
		C1 & $26.08^{+0.05}_{-0.06}$ & 0                       & 0                          & $ 1.769^{+0.005}_{-0.005}$ & $(7.71^{+0.04}_{-0.04})\times 10^9$\\
		C2 & $ 6.64^{+0.10}_{-0.08}$ & $  -4.72^{+0.05}_{-0.05}$ & $ -2.80^{+0.05}_{-0.07}$ & $  8.24^{+0.13}_{-0.07}$   & $(9.0^{+0.1}_{-0.2})\times 10^7$\\
		C3 & $ 8.07^{+0.07}_{-0.07}$ & $ -13.05^{+0.10}_{-0.10}$ & $  3.00^{+0.10}_{-0.06}$ & $  15.1^{+0.1}_{-0.1}$    & $(3.28^{+0.05}_{-0.06})\times 10^7$\\
		C4 & $ 0.78^{+0.03}_{-0.05}$ & $ -32.9^{+0.6}_{-0.3}$ & $ 16.9^{+0.5}_{-0.5}$ & $ 7.3^{+0.3}_{-1.7}$    & $(1.37^{+4.23}_{-0.09})\times 10^7$\\
		C5 & $ 0.57^{+0.03}_{-0.02}$ & $ -51.3^{+0.5}_{-0.7}$ & $ 4.2^{+1.2}_{-1.0}$ & $ 8.2^{+1.0}_{-1.2}$    & $(7.9^{+4.7}_{-0.1})\times 10^6$\\
		\hline 
	\end{tabular}
	\vspace{0.5em}
\end{table*}

	Importantly, due to the fact that the source appears more compact in the new EVN data (and also in the re-analyzed e-EVN data) than reported previously, also the source brightness temperature is higher than that derived in \citetalias{gabanyi2013}. In particular, from the standard formula\footnotemark
	\begin{eqnarray*}
		T_b &=& \frac{c^2}{2k_B \nu^2} \frac{F}{\pi (\phi/2) ^2 \ln 2}\\
		&=& 9.9 \times 10^{9} \,{\rm K}\times \left( \frac{\nu}{\rm 1.6 \,GHz} \right)^{-2}\\
		&&
		\times \left( \frac{F}{\rm 10\, mJy}\right)\left( \frac{\phi_{\rm maj}}{\rm 1 \, mas}\right)\left( \frac{\phi_{\rm min}}{\rm 1 \, m as}\right),
	\end{eqnarray*}
	where $F$, $\nu$, and $\phi_{\rm maj}$/$\phi_{\rm min}$ are the flux density, observation frequency, and the major-/minor-axis size of the beam or of the Gaussian component, respectively \citep[e.g.][]{akiyama2015}, we find the peak brightness temperature of $1.8 \times 10^9$~K for the peak intensity of the image 19.9~mJy~beam$^{-1}$. Note that since the core component is only marginally resolved, as shown in Fig.~\ref{fig:image} (see also Table~\ref{tab:cgauss}), the corresponding uncertainties in the derived value of $T_b$ are relatively large; still, the peak value of $1.8 \times 10^9$~K corresponds to the safe \emph{lower} limit for the source brightness temperature.
	\footnote[5]{Here we ignore the redshift-correction factor $(1+z)$, due to the anticipated source distance of $\sim 600$~Mpc only (following \citetalias{tanaka2014}), noting that this particular choice does not affect the main results or conclusions presented in this Letter.}
	
	 Previous surveys of BL Lacs found typical peak intensities of a few to 100~mJy~beam$^{-1}$ at 8~GHz \citep[][hereafter \citetalias{liuzzo2013}]{liuzzo2013} and $\sim$~10-100~mJy beam$^{-1}$ at 1.6~GHz \citep{giroletti2006}, corresponding to the brightness temperatures of $\sim10^{8-10}$~K. Recently, \citet{piner2014} reported typical peak brightness temperatures of $\sim10^{9-10}$~K for the analyzed sample of 20 BL Lac objects detected at TeV photon energies. The peak intensity and peak brightness temperature derived here for J1943+213 are therefore consistent with those characterizing BL Lacs in general, and TeV BL Lacs in particular.
	
	On the other hand, the other components in the J1943+213 jet are fairly resolved, with the corresponding brightness temperature of $\sim 10^{6-8}$~K, again in agreement to what is observed in other BL Lac objects \citep{giroletti2006,piner2014}. We also note that the brightness temperatures of all our Gaussian components are anyway much larger than the brightness temperatures of $\sim10^{3-4}$~K and $\sim10^{1-2}$~K at 1.4~GHz measured for the paradigmatic PWNe Crab \citep{bietenholz1991} and 3C~58 \citep{bietenholz2006}.
	
	\section{Discussion}
	
	Our new observations provide a very strong, basically even conclusive evidence for the blazar identification of the enigmatic TeV emitter J1943+213: the milliarcsecond-scale structure of the source displays a core--jet morphology and a high brightness temperature, both of which are characteristic of HBLs as a class (and at the same time inconsistent with radio properties of Galactic PWNe).
		
	\subsection{J1943+213 as an extreme HBL}
	Extreme HBLs form a peculiar sub-class of BL Lac objects, with intrinsic (i.e., corrected for the absorption on the extragalactic background light) TeV spectra equivalent to, or even harder than $F(\nu) \propto \nu^{-1}$. Such spectra, in the framework of conventional one-zone synchrotron-self-Compton (SSC) models widely applied to broad-band SEDs of TeV BL Lacs \citep[e.g.][]{kino2002}, requires rather extreme physical conditions within blazar emitting zones, and in particular very high jet bulk Lorentz factors, $\Gamma \gtrsim 30$, weak jet magnetic fields, $B' \sim 1-10$~mG, and very high minimum energies of ultra-relativistic jet electrons, $\gamma_{\rm min} \geq 1000$ \citep[see, e.g.,][]{katarzynski2006,tavecchio2009,tavecchio2010}. Indeed, $\Gamma \simeq 70$, $B' \simeq 0.8$\,mG, and $\gamma_{\rm min}=10^5$ have been estimated by \citetalias{tanaka2014} from the one-zone SSC modeling of the broad-band spectrum of J1943+213. We note that up to now only about 10 extreme HBLs are known.
	
	Our observations reveal also a very low ``X-ray defined radio-loudness'' of J1943+213, which is however comparable to those characterizing other extreme HBLs, with typical X-ray-to-radio flux ratios of, roughly, $F_{\rm x}(0.1-2.4\,{\rm keV})/F_{\rm radio}(1.4\,{\rm GHz})\gtrsim10^{4}$ \citep[e.g.][]{bonnoli2015}. We measure the J1943+213 core flux spectral energy density of $\sim$26.1~mJy, which corresponds to the monochromatic energy flux of $4.3\times 10^{-16}$~erg~cm$^{-2}$~s$^{-1}$. Meanwhile, the {\it Suzaku} 0.5--25~keV spectrum of the source (\citetalias{tanaka2014}) extrapolated down to lower photon energies, returns $F_{\rm x}(0.1-2.4\,{\rm keV})\sim 3.7\times 10^{-11}$~erg~cm$^{-2}$~s$^{-1}$. These result in $F_{\rm x}(0.1-2.4\,{\rm keV})/F_{\rm radio}(1.6\,{\rm GHz})\sim 9\times10^{4}$, providing yet another evidence in support of the blazar identification of J1943+213.
	
	Two-epoch VLBI observations with EVN do not show significant flux variations, which might be consistent with previous X-ray and TeV $\gamma$-ray observations showing no time variability at X-ray and TeV bands \citep{shahinyan2015}. Although the statistical errors of the total flux are small ($1\sigma$ errors are 0.09~mJy and 0.02~mJy for 2011 and 2015 data, respectively), if we adopt typical {\it a-priori} calibration errors of $\lesssim 10$~\% for EVN \citep[e.g.][]{bondi1994}, the total flux densities of $42\pm4$~mJy in 2014 and 34$\pm 3$~mJy in 2011 are mostly consistent within $3\sigma$. However, a flux density of $22.4\pm0.3$~mJy was recently reported on e-MERLIN observations at 1.5~GHz in 2013 December and 2014 June \citep{gabanyi2015}, which would be an evidence for the flux variation combined with our data \citep{straal2016}. Since all of four VLBI observations were performed with quite different $uv$-coverages and arrays, more homogeneous observations will be necessary to conclusively confirm the flux variations.
	
	We remark that, since the source is close to the Galactic plane, the flux variation in radio regimes can be \emph{extrinsic} due to the interstellar scintillation \citep[e.g.][]{akiyama2016}. For instance, the refractive scale of the interstellar scintillation at the source region is 2.81~mas at 1.6 GHz estimated from the NE2001 model\footnote{\url{http://www.nrl.navy.mil/rsd/RORF/ne2001/}} \citep{cordes2002,cordes2003}, which is comparable to the FWHM size of the radio core. This suggests that a source with the angular size of $\lesssim 2.81$~mas can be modulated by refractive interstellar scattering, and therefore the radio core can be extrinsically scintillated \citep[see][for details]{akiyama2016}.
	
	\subsection{The source compactness and core dominance}
	It is instructive to evaluate the source compactness (SC; e.g. \citetalias{liuzzo2013}) and the core dominance \citep[CD;][]{liuzzo2009} of J1943+213, for a direct comparison with other BL Lacs. The SC is defined as a ratio of total flux densities (or powers) measured on milliarcsecond and arcsecond scales. On the other hand, the CD is the ratio between the observed core radio power and the expected core radio power estimated from the umbeamed total radio power at lower frequencies, based on the relation given in \cite{giovannini1994}. The CD is a good proxy of the jet Doppler factor for blazar sources. 

	The total flux density of J1943+213 on arcsecond scales at 1.4~GHz, as measured with the NRAO VLA Sky Survey (NVSS), is $102.6 \pm3.6$~mJy, giving the SC of $ \sim 0.4$. Adopting the source distance of 600~Mpc, as advocated in \citetalias{tanaka2014}, the total radio power of J1943+213 on milliarcsecond and arcsecond scales reads as $\sim 1.8\times  10^{24}$~W~Hz$^{-1}$ and $\sim 4.4\times 10^{24}$~W~Hz$^{-1}$, respectively. The unbeamed core radio power is estimated to be $\sim 5.8 \times 10^{22}$~W~Hz$^{-1}$, assuming the average spectrum index of 0.7 between 408~MHz and 1.4~GHz, following \citetalias{liuzzo2013}. This gives the CD of 30. 

	The estimated SC and CD are in the range of typical values found for blazars, where the VLBI scale core flux density is strongly enhanced by the Doppler beaming \citepalias{liuzzo2013}. Moreover, they are qualitatively consistent with the large Doppler factor inffered from the one-zone SSC model applied by \citetalias{tanaka2014} to the SED of J1943+213. This again supports the blazar identification of the source, and implies that there is nothing unusual about it \emph{as long as small-scale radio properties are considered}.

\subsection{A new population of unrecognized BL Lacs?}
		
J1943+213 has been detected with H.E.S.S. at TeV photon energies \emph{only} because it happened to be located close to the Galactic plane, and as such it was covered by the H.E.S.S. Galactic Plane Survey. Note in this context that till now no systematic TeV survey has been carried out at high Galactic latitudes by the currently operating Cherenkov telescopes. At the same time, the source is very weak in the GeV range, so that it could be detected (at a marginal level and in a limited energy range) only after five years of the accumulation of the all-sky {\it Fermi}/LAT data. The serendipitous H.E.S.S. discovery has triggered the follow-up X-ray \citepalias{tanaka2014}, near-infrared \citep{peter2014}, and high-resolution radio (\citetalias{gabanyi2013}, this Letter) observations, which enabled to identify the source as a blazar of the ``extreme HBL'' type. Remarkably, at optical and infrared frequencies, J1943+213 lacks any signatures of an AGN activity. In other words, if not the particular location of J1943+213 on the sky, this object would never be recognized, targeted, and classified as a blazar the first place.

The question is if J1943+213 is an anomaly/outliers in the diverse blazar population, or if it represents possibly quite numerous population of previously unrecognized, particularly extreme HBLs. Since H.E.S.S. found this source over the 1224~deg$^2$ survey \citep{carrigan2013}, one can expect $\sim34$ sources detectable with a sensitivity of existing Chelenkov telescopes over the whole sky, which are already three times larger than the known population of extreme HBLs. This relevant question can be addressed in the near future only by performing a systematic survey of nearby and seemingly ``quiet'' elliptical galaxies with hard X-ray satellites (NuSTAR, ASTRO-H), high-resolution radio interferometers (VLBI), and sensitive TeV telescopes (Cherenkov Telescope Array).

	\section{Conclusion}
	Here we present the new EVN observations of HESS~J1943+213 at 1.6~GHz. Our results provide a very strong, basically even conclusive evidence for the blazar identification of this enigmatic TeV emitter:
	\begin{enumerate}
		\item Our high-resolution radio maps reveal that the milliarcsecond-scale structure of J1943+213 displays a clear core--jet morphology, which is typically seen in low-luminosity blazars. The brightness temperatures of the core and other small-scale jet components are consistent with those established for the TeV-emitting BL Lacs as well, and at the same time inconsistent with radio properties of Galactic PWNe.
		\item Careful re-analysis of previous e-EVN observations performed in 2011 \citepalias{gabanyi2013} provides a robust fringe detections on long baselines. The obtained source structure is broadly consistent with our new observations. 
		\item The radio source compactness, radio core dominance and the X-ray--to--radio flux ratio characterizing J1943+213, are all in agreement to what is observed in other extreme HBLs, and are indicative of a significant Doppler beaming shaping radio appearance of the source on milliarcsecond scales.
	\end{enumerate}
	As such, J1943+213 --- which lack any optical/infrared signatures of an AGN activity --- may be either an outliers in the diverse blazar population, or a member of a numerous population of previously unrecognized, particularly extreme HBLs. Regardless of this, our and \citetalias{gabanyi2013} results demonstrate that the low-frequency VLBI observations with high-angular resolution are in general indispensable for a proper identification of extragalactic $\gamma$-ray emitters.		
	
	\acknowledgements
	We thank an anonymous referee for his/her useful suggestions to improve this Letter. 
	This work is supported by JSPS Postdoctoral Fellowships for Research Abroad and grants from the National Science Foundation (NSF). M.H. and K.A. are financially supported by the MEXT/JSPS KAKENHI Grant Numbers 24540242, 25120007 and 25120008. We acknowledge a contribution from the Italian Foreign Affair Minister under the bilateral scientific collaboration between Italy and Japan. {\L}.S. was supported by Polish NSC grant DEC-2012/04/A/ST9/00083.
	
	The European VLBI Network is a joint facility of independent European, African, Asian, and North American radio astronomy institutes. Scientific results from data presented in this publication are derived from the following EVN project code(s): RSG03 and EA056.
	
	\facility{EVN}

\end{document}